\documentclass[twocolumn]{aastex62}

\def\ltsima{$\; \buildrel < \over \sim\;$}
\def\ltsim{\lower.5ex\hbox{\ltsima}}
\def\gtsima{$\; \buildrel > \over\sim \;$}
\def\gtsim{\lower.5ex\hbox{\gtsima}}
\def\ms{$M_{\odot}$ }
\def\msp{$M_{\odot}$}

\received{July 17, 2018}
\revised{xxx}
\accepted{August 5, 2018}

\shorttitle{Chemical evolution of Zn driven by magnetorotational supernovae}
\shortauthors{Tsujimoto \& Nishimura}

\begin{document}

\title{Early chemical evolution of Zn driven by magnetorotational supernovae and the pathway to the solar Zn composition}

\correspondingauthor{Takuji Tsujimoto}
\email{taku.tsujimoto@nao.ac.jp}

\author[0000-0002-9397-3658]{Takuji Tsujimoto}
\affil{National Astronomical Observatory of Japan, Mitaka, Tokyo 181-8588, Japan}

\author[0000-0002-0842-7856]{Nobuya Nishimura}
\affiliation{Yukawa Institute for Theoretical Physics, Kyoto University, Kyoto 606-8502, Japan}



\begin{abstract}

The site of Zn production remains an elusive and challenging problem in astrophysics. A large enhancement of the [Zn/Fe] ratios of very metal-poor stars in the Galactic halo suggests the death of short-lived massive stars, i.e., core-collapse supernovae (CCSNe), as one major site for Zn production. Previous studies have claimed that some specific CCSNe can produce Zn in sufficient quantities. However, it remains unclear which models can withstand the critical test of observations. Using a Zn abundance feature similar to that of r-process elements in faint satellite galaxies, we find evidence that Zn production took place through much rarer events than canonical CCSNe. This finding can be unified with the implied decrease in the rate of Zn production with an increasing metallicity for Galactic halo stars, which narrows down the major site of Zn production in the early galaxy to magneto-rotational SNe (MR-SNe). On the other hand, in the later phase of galactic evolution, we predict that the major Zn-production site switched from MR-SNe to thermonuclear SNe (SNe Ia). According to this scenario, an accumulation of the contributions from two types of SNe eventually led to the solar isotope composition of Zn which mainly owes $^{66,68}$Zn to MR-SNe and $^{64}$Zn to SNe Ia triggered by He-detonation. The requirement of Zn production in SNe Ia sheds a new light on the hot debate on the scenario for SN Ia progenitors, suggesting that a He-detonation model might be one major channel for SNe Ia.

\end{abstract}

\keywords{nuclear reactions, nucleosynthesis, abundances --- Galaxy: evolution --- galaxies: dwarf --- galaxies: evolution --- Local Group --- stars: abundances}


\section{Introduction}

The recent discovery of gravitational waves from the neutron star merger GW170817 and the subsequent discovery of multi-wavelength electromagnetic counterparts. i.e., kilonova, has remarkably improved our understanding of the origin of r-process elements \citep[e.g.,][]{Smartt_17, Pian_17, Cowperthwaite_17, Thielemann_17}. However, the periodic table is still incomplete in the astrophysical sense in that it includes elements without a clear explanation of their origins. A major example of such an element is a light trans-iron element, Zn. The remarkable feature of Zn abundance is represented by the high [Zn/Fe] abundance ratio, ranging from 0 to +0.6 for very metal-poor stars with [Fe/H]\ltsim $-2.5$ \citep{Cayrel_04}. This implies that Zn was efficiently produced with no time delay at a very early phase of galactic evolution. Therefore, core-collapse supernovae (CCSNe) occurring at the deaths of massive stars should be at least one major site for Zn production. 

However, Zn production through explosive burning in canonical CCSNe has been found insufficient to explain the observed [Zn/Fe] \citep[e.g.,][]{Woosley_95, Thielemann_96}. \citet{Umeda_02} thus proposed hypernova models, i.e., CCSNe with large explosion energies capable of increasing Zn production and lifting the predicted [Zn/Fe] ratios up to the value of +0.4 \citep[see also][]{Tominaga_07}. Another type of CCSNe that has been proposed as a major site of Zn production is electron-capture SNe (ECSNe), which are triggered by the collapse of O+Ne+Mg cores in stars with an initial mass of $\sim8-10$\ms \citep{Wanajo_11, Wanajo_18}. These Zn-production sites were discussed in terms of galactic chemical evolution \citep{Kobayashi_06, Hirai_18}. \citet{Kobayashi_06} found that a high frequency of hypernovae (as much as half of all CCSNe) would be required to achieve the observed broad constancy of [Zn/Fe]$\approx$0 in the range of $-2$\ltsim[Fe/H]\ltsim0 for Galactic stars. The deduced frequency of hypernovae seems extremely high. On the other hand, \citet{Hirai_18} reproduced the observed decreasing trend of [Zn/Fe] in the Sculptor dwarf spheroidal galaxy (dSph), assuming ECSNe to be the major source of Zn. It is, however, questionable whether their models are capable of predicting the non-decreasing feature of [Zn/Fe] in the Galaxy.

Recently, a new channel for producing a large amount of Zn in CCSNe has been proposed. \citet{Nishimura_17} showed that magneto-rotational SNe (MR-SNe) triggered by fast rotations and high magnetic fields \citep[e.g.,][]{Takiwaki_09, Nishimura_15} produce light trans-iron elements including Zn as well as the first- to third-peak r-process isotopes. MR-SNe produce Zn more efficiently than the other two candidates, i.e., as high as $(5-6)\times10^{-3}$\ms for each MR-SN, which is about five times of that for an ECSN \citep{Wanajo_18} or more than ten times that for a hypernova from low-metallicity star \citep{Kobayashi_06}. Such a high Zn-production rate in MR-SNe might be reconciled with the low frequency expected for such events themselves to obtain a consistent view of the chemical enrichment of Zn in galaxies. Therefore, capturing an observational signature indicating whether Zn production is a rare event would be a major clue for understanding the connection of Zn sources to MR-SNe or other type of CCSNe. This issue is of growing significance if we consider the recent claims that canonical CCSNe driven by the neutrino heating are able to give a more important contribution of Zn production than previously thought \citep[e.g.,][]{Harris_17, Curtis_18, Eichler_18}.

The small stellar masses in the Milky Way satellite galaxies offer an advantage for this test. The relatively small number of stars formed in these galaxies allows the detection of rare production events such as neutron star mergers in r-process abundances \citep[][Tsujimoto et al.~2015, 2017]{Tsujimoto_14}.  Figure~1 demonstrates the observed [r-process/H] vs.~[Fe/H] results in the Draco dSph, which suggest that r-process events occurred sporadically around [Fe/H]=$-3$ and $-2.3$ with high r-process yields and did not happen afterwards until the end of metal enrichment. The revealed feature implies a level of rarity of r-process production which may be compatible with the frequency expected for neutron star mergers or MR-SNe \citep{TsujimotoN_15, Tsujimoto_17}. Similar assessment of the frequency of such production events using dSphs can be done for Zn.

\begin{figure}[t]
\vspace{0.4cm}
\begin{center}
\includegraphics[angle=0,width=8cm,clip=true]{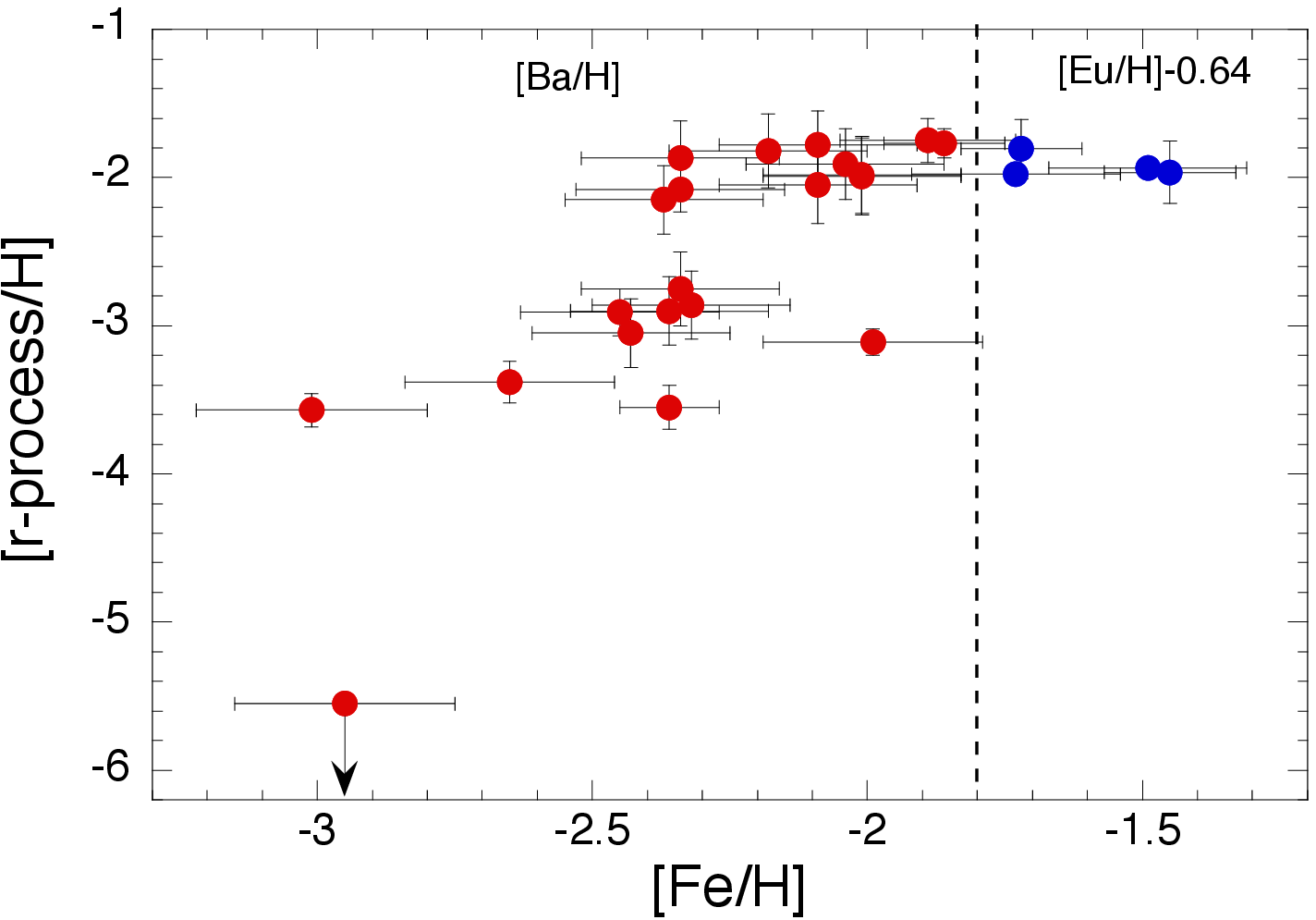}
\end{center}
\caption{Observed r-process abundances vs.~Fe abundances for the Draco dSph. The [Ba/H] data for [Fe/H]$<-1.8$ and the [Eu/H] data for [Fe/H]$>-1.8$ are taken from \citet{Tsujimoto_17} and \citet{Tsujimoto_15}, respectively. The Ba abundances for these low-metallicity stars can be regarded as an r-process Ba abundance based on the pure r-process Ba/Eu ratios measured for several stars. The Eu abundances are shifted by $-0.64$ dex, which corresponds to the value of the solar r-process Ba/Eu ratio, and are thereby equivalent to the r-process Ba abundances for stars with [Fe/H]$>-1.8$.}
\end{figure}

Although such insights are hard to be gained from more massive galaxies such as our own, an assembly of the elemental abundances of solar neighborhood stars, tracing the whole history of Zn enrichment from the first generations of stars to the present, offers a treasure trove of clues as to the identity of the Zn source. In particular, if there is not a sole major site of Zn production (as is the case for many other elements), the evolution of  Zn abundance with the stellar generation will be crucial information. In this Letter, we will begin by narrowing down the possible sites of Zn production based on the set of elemental abundances for dSphs and for the Galaxy. 

\section{Z{\scriptsize n} abundance features in dSphs and in the Galaxy}

\begin{figure}[t]
\vspace{0.4cm}
\begin{center}
\includegraphics[angle=0,width=8cm,clip=true]{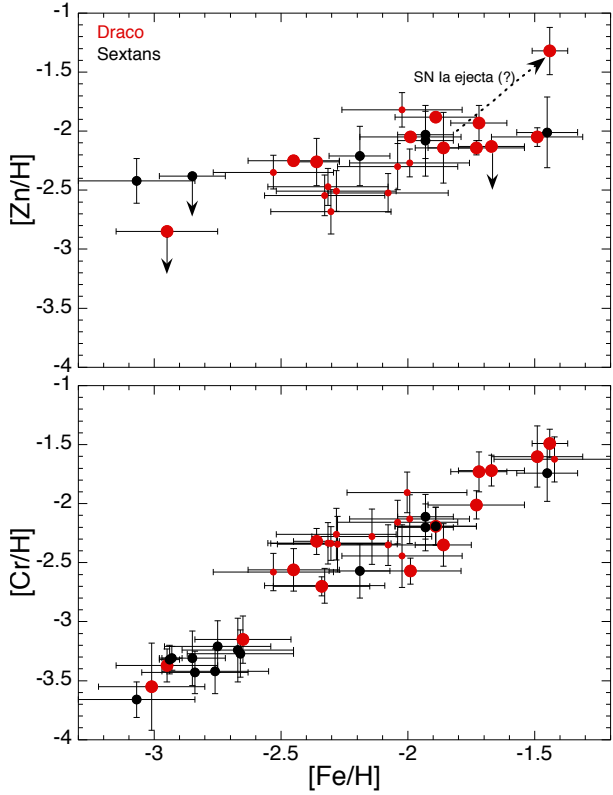}
\end{center}
\caption{Top panel: observed Zn vs.~Fe abundances for the stars in the Draco \citep[red circles,][]{Shetrone_01, Fulbright_04, Cohen_09} and the Sextans \citep[black circles,][]{Shetrone_01, Honda_11} dSphs. For the Draco, the latest Subaru results are attached by small circles (Matsuno et al.~in preparation). There is one Draco star showing a large enhancement of [Zn/H], deviating from those of the other stars. Such an anomalous Zn abundance can be attributed to the unique birth environment, in which interstellar matter is exclusively enriched by the ejecta of an SN Ia (see the text). Bottom panel: same as the top panel, but for Cr.}
\end{figure}

\begin{figure}[t]
\vspace{0.4cm}
\begin{center}
\includegraphics[angle=0,width=8.5cm,clip=true]{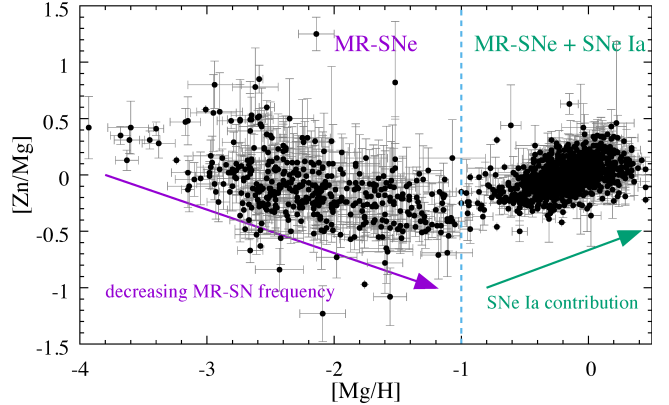}
\end{center}
\caption{Observed correlation of [Zn/Mg] with [Mg/H] for the Galaxy obtained from the SAGA database \citep{Suda_08}. The [Zn/Mg] ratio decreases with increasing [Mg/H] and upturns around [Mg/H]=$-1$. These two discrete trends are attributed to a decrease in the event frequency of MR-SNe as a function of the metallicity and an accumulation of the contribution to Zn from SNe Ia, respectively. As a result, the observed [Zn/Mg] ratio reflects the ratio of contributions from different types of SNe for each element with elapsed time: i.e., MR-SNe to CCSNe for [Mg/H]\ltsim$-1$ and MR-SNe plus SNe Ia to CCSNe for [Mg/H]\gtsim$-1$.}
\end{figure}

Based on the acquired knowledge from r-process abundance feature in dSphs that r-process-production events are sporadic and rare, we begin by examining how Zn enrichment proceeds in dSphs. Figure 2 shows the [Zn/H] versus [Fe/H] diagram (upper panel) for the two classical dSphs, i.e., Draco and Sextans, as compared to the [Cr/H] case (lower panel). These dSphs are faint and have similar stellar masses. It is evident that there is a difference between the  increasing features of Zn and Cr abundances. As is generally expected, [Cr/H] increases in accordance with an increasing [Fe/H],  as both Fe and Cr originate from the same production sites, namely CCSNe and SNe Ia. A similar trend is also seen for other Fe-peak elements (Mn, Ni). By contrast, [Zn/H] shows a subtle increase overall. For [Fe/H]\ltsim$-2$, the Zn abundance is nearly constant with an approximate value of [Zn/H]$\approx -2.4\pm0.1$. Subsequently, [Zn/H] increases only by $\sim 0.3$ dex until [Fe/H]$\approx-1.5$, except for one outlier exhibiting a high [Zn/H] of $\sim -1.3$. This nearly flat evolution, especially the plateau-like feature of Zn abundance for [Fe/H] \ltsim $-2$, suggests that Zn production is characteristic of rarer events than general CCSNe. This rarity contradicts the idea that hypernovae, which account for 50\% of CCSNe, could constitute the major source of Zn production \citep{Kobayashi_06}. Alternatively, the rarity does support MR-SNe as a production source, owing to their low frequency ($\sim$1/100 of CCSNe),  as estimated from the feature of r-process enrichment in the Draco dSph \citep{TsujimotoN_15}.

Since the progenitors of MR-SNe are fast-rotation stars, the emergence of MR-SNe is probably inclined to low-metallicity stars in which the rotational velocity is expected to be high \citep[e.g.,][]{Meynet_06}. Therefore, the efficiency of Zn production in a galaxy is expected to depend upon the metallicity in the sense that a higher Zn production rate is realized at lower metallicity. This prediction is validated based on the observed Zn features of the Galaxy. Figure 3 shows the [Zn/Mg] evolution with respect to [Mg/H] for solar neighborhood stars. [Zn/Mg] exhibits a decreasing trend from [Mg/H]$\sim -4$ to $-1$. This can be interpreted as a result of a decrease in the Zn production rate as a function of metallicity. 

In addition, we clearly see that the [Zn/Mg] trend reverses from a decrease to an increase with increasing [Mg/H] at around [Mg/H]$\sim -1$. This suggests the presence of another site of Zn production during the late stage of galactic evolution which corresponds to the chemical evolution of the Galactic disk. Considering that the timing of the upturn is broadly equivalent to the location of a knee of [$\alpha$/Fe], it is likely that this second site for Zn might be thermonuclear SNe, or type Ia supernovae (SNe Ia). Among the scenarios for SNe Ia, including deflagration and delayed-detonation models \citep[e.g.,][]{Iwamoto_99}, He-detonation models \citep[e.g.,][]{Woosley_11, Pakmor_13} can produce $^{64}$Zn due to high entropies  with strong  $\alpha$-rich freeze-out, similar to those of hypernovae while the deflagration models do not produce any significant amounts of Zn. Indeed, the dominant isotopes synthesized in MR-SNe are neutron-rich $^{66,68}$Zn \citep{Nishimura_17}. Accordingly, we predict that the major composition of solar Zn abundance (i.e.,   $^{64,66,68}$Zn) is the end result of an accumulation of Zn from two different production sites, i.e., MR-SNe and SNe Ia. We note here that ECSNe were recently proposed as major production sites for all isotopes of Zn \citep{Hirai_18}, although this disagrees with the observed features of the [Zn/Mg] versus [Mg/H] diagram. There is another supporting evidence for our interpretation. MR-SNe models predict a large amount of $^{59}$Co together with Zn, while regular CCSNe are unable to produce it sufficiently \citep{Kobayashi_06}. Indeed the observed [Co/Mg] ratio follows an evolutionary path quite similar to that of [Zn/Mg].

Now, let us return to the discussion of the observed features of dSphs, as shown in Figure 2. If we incorporate the abovementioned interpretation of the Zn sites, an increase in [Zn/H] for [Fe/H]\gtsim$-2$ by $\sim$0.3 dex can, at least in part, be attributed to the contribution of Zn from SNe Ia. In fact, there exists a decreasing [Mg/Fe] feature among dSphs beginning at [Fe/H]$\sim -2$, which is thought to be caused by SN Ia enrichment \citep{Tolstoy_09}. Moreover, the identification of one star that exhibits an unusually high Zn abundance ([Zn/H]=$-1.32$) can be discussed within a  framework that considers the presence of SNe Ia. This star does not show an anomalous abundance of r-process elements (Y, Ba) compared to those of the stars at similar [Fe/H]. Interestingly, a quite similar star is found in the ultra-faint galaxy Reticulum II. This star shows [Zn/H]=$-1.27$, a value nearly one order of magnitude greater than those of other stars, although the abundance of r-process isotopes such as Y, Sr, Ba, and Eu are normal \citep{Ji_16}. These two identical stars share low [$\alpha$/Fe] ratios such as [Mg/Fe]=$-0.19$ (Draco) and $-0.08$ (Ret II). All elemental features combined suggest that the two stars originated in gas that was exclusively enriched by an SN Ia. Such anomalous Zn stars are also seen in the Galactic halo, namely for G4-36 and CS 22966-043 \citep{Ivans_03,Tsujimoto_04}.  

\section{Chemical evolution of the Galaxy}

We try to reproduce the overall evolution of Zn abundance in the Galaxy using one-zone chemical evolution models to test the hypothesis that the major Zn sites are MR-SNe and SNe Ia. For simplicity, we ignore the potential contributions to Zn enrichment from other types of SNe, such as regular CCSNe, hypernovae, and ECSNe. For the Zn yield from each MR-SN, we adopt $5\times10^{-3}$\msp, which is a standard value in models with high neutrino luminosities  \citep{Nishimura_17}. It should be noted that the models for regular CCSNe with low-metallicities such as [Fe/H]=$-1.3$ are predicted to produce less Zn than MR-SNe models by about two orders of magnitude \citep{Nomoto_13}. The critical parameter in our models is the dependence of the MR-SN frequency upon the metallicity. In our calculations, we define $f_{\rm MRSN}$ as the number fraction of MR-SNe against CCSNe for each generation of stars formed and determine it so as to reproduce the observed trend under the condition that $f_{\rm MRSN}$ decreases in accordance with increasing metallicity. Here we apply a higher fraction of fast-rotating stars which end with MR-SNe at lower metallicities \citep[see][]{Nishimura_17}. The adopted $f_{\rm MRSN}$ is found to be 50\% for [Fe/H]$\leq -3.3$, 5\% for $-3.3<$[Fe/H]$\leq-2.5$, 3\% for $-2.5<$[Fe/H]$\leq-0.6$, and 0.5\% for [Fe/H]$\geq-0.6$. A few percent as $f_{\rm MRSN}$ for [Fe/H]\gtsim$-3$ is also capable of explaining an increase in [Zn/H] from $-2.5$ (or $-3$) to $-2$ in the dShs in Figure~2.

For SNe Ia, we assume the Zn yield to be $2\times10^{-3}$\ms so as to adjust the enrichment level of Zn to match the stellar abundances of disk stars. The adopted yield is broadly consistent with those predicted by some He-detonation models \citep{Woosley_11}. In addition, as a supporting argument, the predicted [Zn/H] value within ejecta of SN Ia that have swept up interstellar matter with [Zn/H]=$-2$ is estimated to be [Zn/H]$\sim-1.5$, which is compatible within error with that of a Draco star showing an anomalous Zn abundance. For the delay-time distribution (DTD) of SNe Ia, we adopt DTD $\propto t_{\rm delay}^{-1}$ within the range $0.1\leq t_{ \rm delay}\leq10$ Gyr. 

\begin{figure}[t]
\vspace{0.4cm}
\begin{center}
\includegraphics[angle=0,width=8.5cm,clip=true]{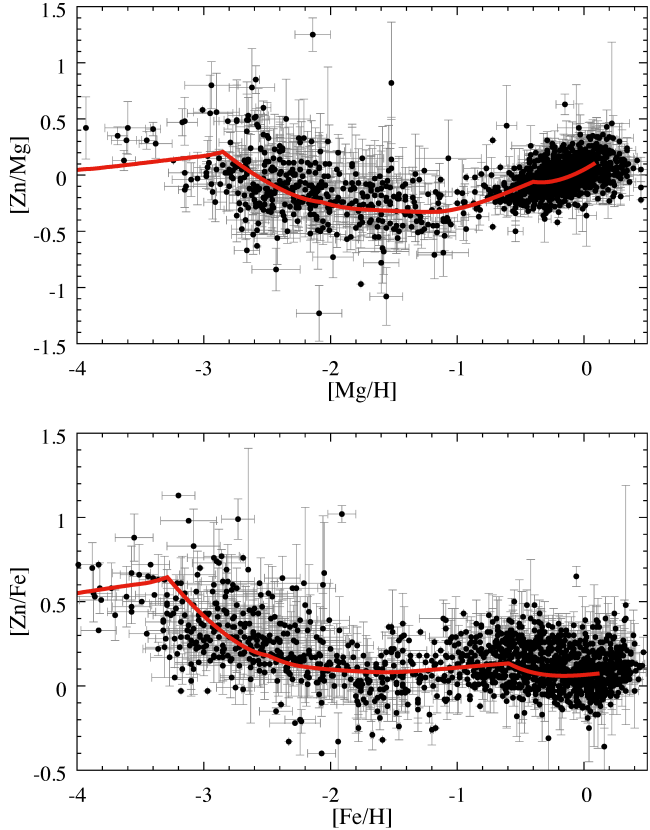}
\end{center}
\caption{Top panel: predicted [Zn/Mg] evolution against [Mg/H] for the Galaxy, as compared with observations. Bottom panel: same as the top panel but against Fe.}
\end{figure}

To compare theoretical results with the observational data of solar neighborhood stars, we use a thick disk model since thick disk stars cover a wide range of [Fe/H] (namely $-3$\ltsim [Fe/H]\ltsim+0.3; Ruchti et al. 2011), comparable to that in the solar neighborhood, and there is only a little difference in [Zn/Fe] ratio from those of thin disk stars \citep{Bensby_05}. In addition, the rapid formation of a thick disk enables its early stage to be assigned to the phase of halo formation. We assume that the thick disk was formed by infalling gas from outside of the disk region. Taking into account the relatively rapid formation of the thick disk, we adopt a rather short infall timescale of 0.5 Gyr. The star formation rate is assumed to be proportional to the gas fraction with a constant coefficient of 1 Gyr$^{-1}$ for a duration of 2.5 Gyr. Details are described in \citet{Tsujimoto_12}.

We calculate the evolution of Zn abundance with respect to both Mg and Fe abundances. The results are shown in two panels of Figure 4. We find good agreement between our model results and observations. From the bottom of [Zn/Mg]$\sim -0.3$, a mean ratio of [Zn/Mg] increases to 0 at around solar metallicity. Since the increment of $\sim$0.3 dex is caused by Zn released from SNe Ia, it turns out that our model's solar Zn abundance comprises the nearly equal contributions from MR-SNe and SNe Ia. This is consistent with the similar percentage of solar Zn isotopes between $^{64}$Zn (48.6\%) and $^{66}$Zn+$^{68}$Zn (46.7\%). For [Fe/H]\gtsim$-0.6$, we see the path with less Zn production predicted by a very low $f_{\rm MRSN}$, which may, alternatively, imply the possibility of the Zn yield in He-detonation SN Ia giving a smaller amount for metal-richer stars.

In our models, we assume that MR-SN events are not rare for a very low-metallicity range such as [Fe/H] $<-3$. This hypothesis seems like the only solution to give a mean value of [Zn/Fe]$\sim +0.5$. A high frequency of MR-SNe in the early phase may be compatible with the lack of supporting evidence for the delayed r-process enrichment in the Galactic halo \citep{Roederer_13}. Stars with high [Zn/Fe] ratios of $\sim +0.5$ for [Fe/H]$<-3$ are also identified in dSphs such as the Ursa Minor and the Sculptor \citep{Skuladottir_18}, in addition to one Sextans star shown in Figure~2.

\section{Discussions}

We propose that MR-SNe are the major sites of Zn production during the early epoch of galactic evolution. However, this does not necessarily mean that MR-SNe simultaneously drive enrichment of heavy r-process elements such as Ba and Eu since models with high Zn production are associated with a low production efficiency of heavy r-process elements: CCSNe with stronger, but still intermediate magnetic fields produce Fe and Zn while ones with much stronger magnetic fields produce r-process elements \citep{Nishimura_17}. On the other hand, we see a similarity in the evolutionary paths of abundances between Zn and r-process elements in dSphs (compare Fig.~1 and the top panel of Fig.~2). More research on both the observational and theoretical sides will be needed to clarify this connection.

Regular CCSNe driven by the neutrino heating are proposed to be capable of making Zn/Fe like a solar ratio \citep{Frohlich_06a, Curtis_18}. Their models yield a better fit to the Fe-group and also produce nuclei up to $A$=100 \citep{Frohlich_06b, Eichler_18}. These studies imply that regular CCSNe can be one major site of Zn production, at least in the range of $-2$\ltsim[Fe/H]\ltsim$-1$. However, we show that the observed abundance features of dSphs and the Galaxy argue against the continuous Zn production at the same rate over galactic evolution as predicted by their models. Our study poses an open question to the nucleosynthesis models for regular CCSNe.

We suggest that the upturn of Zn/Fe (Zn/Mg) at very low metallicities could be related to less mass loss and thus less loss of  angular momentum, leading to a higher fraction of fast-rotating objects which end with CCSNe with stronger magnetic fields, i.e., MR-SNe. In this view, a high fraction of MR-SNe at very low metallicities would make few stars with low Zn/Fe ratios ([Zn/Fe]$<$0), which leads to a relatively smaller scatter in [Zn/Fe] compared to that of [r-process/Fe], in spite of the presence of the rarity of Zn production event. On the other hand, there is a different argument that this trend reflects the abundance ratios predicted by hypernova models with different progenitor masses \citep{Tominaga_07}. However, in this case, a Zn contributor different from SNe Ia in the later phase is required to avoid an overproduction of $^{64}$Zn in the solar composition. This issue can be also discussed in terms of the upper side of scatter in [Zn/Fe]. The nucleosynthesis Zn/Fe ratios predicted by individual MR-SN models are distributed up to [Zn/Fe]$\sim +1.5$. This may promise that a large contribution to Zn enrichment from MR-SNe in the early Galaxy satisfactorily explains an observed scatter attaining above [Zn/Fe]$\sim +1$ among very low-metallicity stars, while hypernova models predict [Zn/Fe]$<+0.7$. 

Together with MR-SNe, SNe Ia triggered by He detonations are proposed as another site of Zn production in this study. The He-detonation model can be realized in both of two well-known scenarios for SNe Ia: a single degenerate scenario \citep[an accreting white dwarf from a non-degenerate companion star:][]{Woosley_11} and a double degenerate scenario \citep[a merger of double white dwarfs:][]{Pakmor_13}. Recently, this He-detonation mode of the explosion has been highlighted as a mechanism to explain the observed early emission feature for SNe Ia \citep{Jiang_17, Noebauer_17}. Other observational indications also ask strongly for a He-detonation contribution to SNe Ia's \citep{Shen_18}. In terms of nucleosynthesis, we claim that the presence of this pathway is important to SN Ia explosions. On the other hand, from Mn observations \citep[e.g.,][]{Mishenina_15}, we see an increase in [Mn/Fe] above [Fe/H]=$-1$. This argues for a single-degenerate deflagration/detonation contribution to SN Ia's, which have high central densities and electron captures reducing Y$_e$.  Accordingly, combining the Mn observations arguing for deflagration/detonation type and the Zn observations arguing for He-detonation type, this gives a nice clue on the combination of both types, as recent observational surveys indicate \citep[e.g.,][]{Livio_18}.

\acknowledgements

We thank Friedrich-K. Thielemann for helpful suggestions and comments on the manuscript. This work was supported by JSPS KAKENHI Grant Numbers 18H01258, 18H04593, and 16H04081 and by MEXT Japan (Priority Issue on Post K Computer: Elucidation of the fundamental laws and evolution of the universe).

\end{document}